\documentclass[runningheads]{llncs}
\usepackage[T1]{fontenc}
\usepackage{graphicx,verbatim}
\usepackage{booktabs} 
\usepackage{tabularx}
\newcolumntype{Y}{>{\centering\arraybackslash}X}
\usepackage{float}
\usepackage[labelformat=simple]{subcaption}
\usepackage{adjustbox}
\usepackage{microtype}
\usepackage{bm}
\usepackage{pifont}
\usepackage{graphicx}
\usepackage[hidelinks]{hyperref}
\usepackage{color}

\usepackage{amssymb}
\usepackage{amsmath}
\usepackage{dsfont}

\begin{document}
\title{RegScore: Scoring Systems for Regression Tasks}
\author{Michal K. Grzeszczyk\inst{1,2} \and
Tomasz Szczepański \inst{1} \and
Pawel Renc\inst{2,3} \and
Siyeop Yoon \inst{2} \and
Jerome Charton \inst{2} \and
Tomasz Trzciński \inst{4,5} \and
Arkadiusz Sitek \inst{2} 
}
\authorrunning{M.K. Grzeszczyk et al.}
%
\institute{Sano Centre for Computational Medicine, Cracow, Poland 
\email{m.grzeszczyk@sanoscience.org}\\\and
Massachusetts General Hospital, Harvard Medical School, Boston, MA, USA\\ \and
AGH University of Krakow, Cracow, Poland \\ \and
Warsaw University of Technology, Warsaw, Poland \\ \and
Research Institute IDEAS, Warsaw, Poland
}

\maketitle              
\begin{abstract}
Scoring systems are widely adopted in medical applications for their inherent simplicity and transparency, particularly for classification tasks involving tabular data. In this work, we introduce RegScore, a novel, sparse, and interpretable scoring system specifically designed for regression tasks. Unlike conventional scoring systems constrained to integer-valued coefficients, RegScore leverages beam search and k-sparse ridge regression to relax these restrictions, thus enhancing predictive performance. We extend RegScore to bimodal deep learning by integrating tabular data with medical images. We utilize the classification token from the TIP (Tabular Image Pretraining) transformer to generate Personalized Linear Regression parameters and a Personalized RegScore, enabling individualized scoring. We demonstrate the effectiveness of RegScore by estimating mean Pulmonary Artery Pressure using tabular data and further refine these estimates by incorporating cardiac MRI images. Experimental results show that RegScore and its personalized bimodal extensions achieve performance comparable to, or better than, state-of-the-art black-box models. Our method provides a transparent and interpretable approach for regression tasks in clinical settings, promoting more informed and trustworthy decision-making.
We provide our code at \url{https://github.com/SanoScience/RegScore}.


\end{abstract}

\section{Introduction}
Scoring systems are sparse linear models that require the addition of points conditioned on binary features that sum to the final score. For example, in the CHADS$_{2}$ \cite{gage2001validation} system, if the age of the patient is higher or equal 75 ($age\geq75$ binary feature), 1 point is added to the final score of stroke risk. Based on the sum of points, a probability can be derived from the pre-computed table or a non-linear function. There are a large number of scoring systems in healthcare, such as CHADS$_{2}$ or NEWS$_{2}$ \cite{SMITH2019260} as clinicians tend to favor methods that are easier to use and interpret, even if they are less accurate than deep learning models. The most popular approaches for data-driven scoring systems creation involve training penalized logistic regression (LR). Then, the coefficients are rounded or have $\pm$1 assigned depending on their sign as the Unit method in \cite{burgess1928factors}. Ustun and Rudin introduced the Risk Supersparse Linear Integer Model (RiskSLIM) \cite{ustun2019learning} whose discrete coefficients are found with Integer Programming. Based on this approach, Multiclass Interpretable Scoring Systems (MISS) expanded scoring system use beyond two classes \cite{grzeszczyk2024miss}. Liu \textit{et al.} \cite{liu2022fasterrisk} presented FasterRisk, which finds scoring systems in a three-step process of solving sparse logistic regression via beam search (BS), finding a pool of nearly optimal solutions with continuous coefficients and rounding them. Although interpretable, traditional scoring systems may fall short when diagnosing conditions defined by thresholding a continuous variable. For example, Pulmonary Hypertension (PH) is diagnosed when the invasively measured mean Pulmonary Artery Pressure (mPAP) exceeds 20 mmHg \cite{MHoeper2017}. Instead of assigning arbitrary points, a more informative approach would be to develop scoring systems that reflect the relationship between features and mPAP.

Furthermore, diagnostics in modern healthcare involve collecting multimodal data in the form of images and tabular records. This aspect led to the development of methods that allow the injection of clinical data into vision deep learning models. Modules like Dynamic Affine Feature Map Transform (DAFT) \cite{polsterl2021combining}, TabAttention \cite{grzeszczyk2023tabattention} or TabMixer \cite{grzeszczyk2024tabmixer} enhance the interaction between imaging and tabular data via affine transformations, attention learning conditioned on tabular data or multilayer perceptron-based mixing of multimodal features. Such methods have surpassed na\"{i}ve approaches for merging both modalities based on concatenation \cite{spasov2019parameter}, maximum value selection \cite{vale2021long} or multiplication \cite{duanmu2020prediction}. Tabular data can also improve unimodal models when used during self-supervised learning (SSL) \cite{hager2023best} or for guiding image feature learning \cite{jiangtabular}. The SSL on both modalities and bimodal inference in the Tabular Image Pretraining (TIP) achieves state-of-the-art results for imaging and tabular data \cite{du2024tip}. Unfortunately, all these approaches are weakly explainable and do not take into account the interpretability of tabular features. Even though one can analyze feature attribution \cite{hager2023best} or attention scores of the classification (CLS) token - serving as a learned representation for classification in transformers \cite{du2024tip} - these methods offer only limited interpretability.

In this paper, we present RegScore, a sparse, interpretable, transparent scoring system for regression tasks. By relaxing integer-only constraints of points in scoring systems and changing the task from classification to regression, we show two ways to create RegScore. Firstly, we find the solution to a sparse ridge regression problem on binary features based on BS \cite{liu2022fasterrisk}. The second approach is to solve it with OKRidge (OKR) \cite{NEURIPS2023_80f48ffa}. Further, we leverage the interpretability of tabular data and show how to produce interpretable predictions for bimodal deep learning models. Given the CLS token from the TIP transformer, instead of computing the final output, we generate weights for linear regression computed with tabular features. We dub this approach Personalized Linear Regression (PLR) since separate linear operations are performed for each of the samples. Similarly, we dynamically mask binary features to produce a Personalized RegScore (PRS) from the CLS token. Our contributions are as follows: (1) we introduce RegScore, a scoring system for regression tasks, (2) we present PLR and PRS, two approaches for generating interpretable predictions from bimodal deep learning architecture, which while being more interpretable are competitive to other solutions, and (3) we apply the presented methods to the task of PH diagnosis and show that RegScore can outperform classification scoring systems by a significant margin.

\section{Method}

\begin{figure}[t!]
    \centering
    \includegraphics[width=\textwidth]{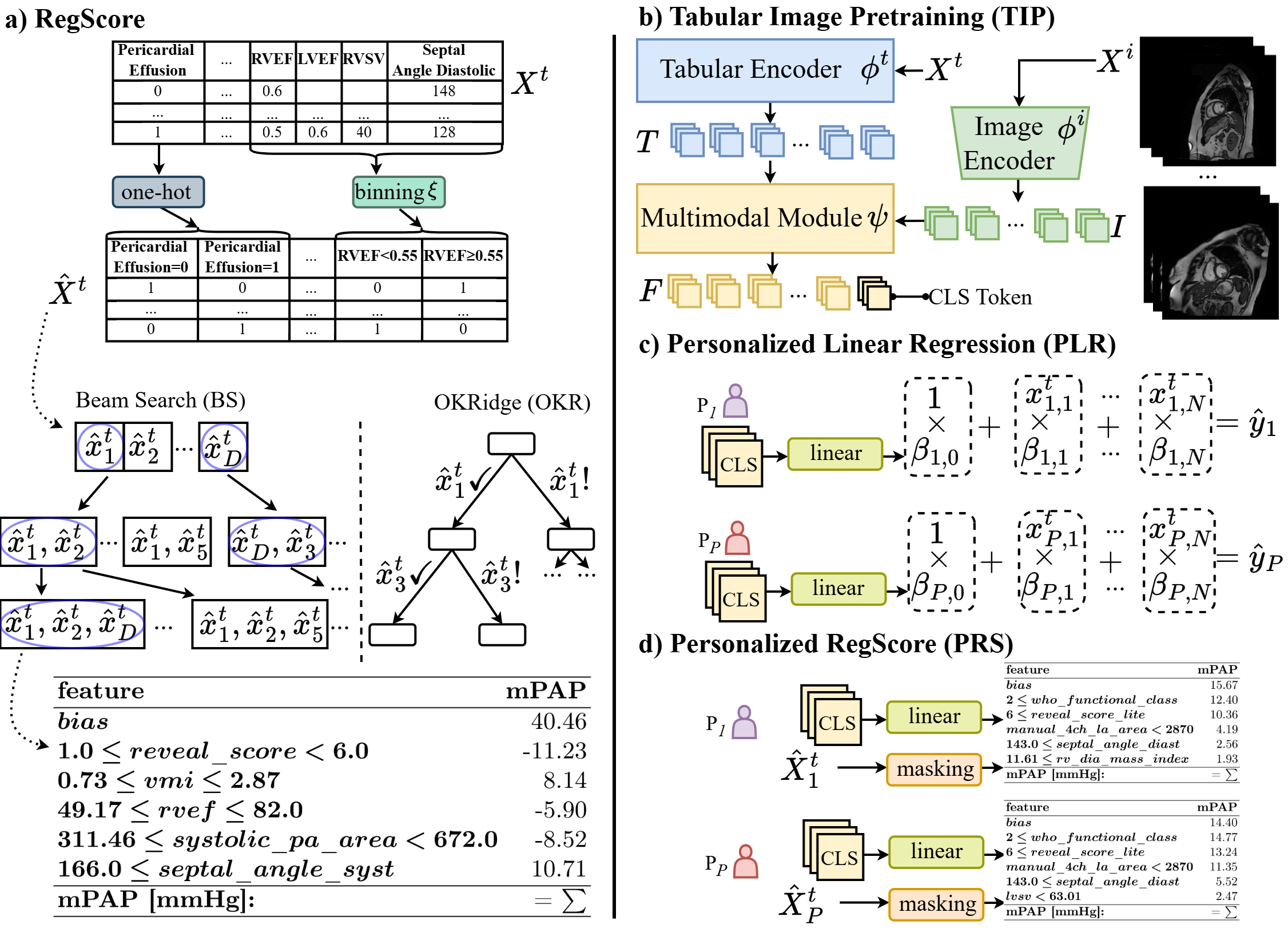}
    \caption{Given the tabular ($X^t$) dataset, we produce the discretized set ($\hat{X}^t$) to create RegScore with beam search or OKRidge (a). We further leverage the TIP method \cite{du2024tip} with imaging data ($X^i$) (b) to compute CLS token for every sample and create interpretable regression predictions with PLR (c) and PRS (d).}
    \label{fig:overview}
\end{figure}

In this section, we present two methods for generating RegScore from tabular data, followed by a description of how the CLS token from the TIP transformer is used to derive PLR and PRS. An overview of our approach is shown in Fig. \ref{fig:overview}.

Let $ (X^t = [x^t_1, \ldots, x^t_N] \in \mathbb{R}^{P\times N}, X^i \in \mathbb{R}^{P \times H \times W \times 3}) $ be a dataset consisting of tabular-image pairs, where \( P \) is the number of samples and \( N \) is the number of features. We construct a binarized set $(\hat{X}^{t} = [\hat{x}^t_1, \ldots, \hat{x}^t_D] \in \{0,1\}^{P\times D})$ by one-hot encoding $N_{cat}$ categorical features and discretizing $N-N_{cat}$ continuous features using a discretization function $\xi$. Various implementations of $\xi$ exist; in this work, we consider the Minimum Description Length Principle (MDLP) \cite{fayyad1993multi} and tertile binning. Given $\hat{X}^t$, we generate RegScore.

\noindent \textbf{RegScore.} The objective of RegScore is to minimize the mean squared loss under a sparsity constraint, resulting in a sparse ridge regression problem:

\begin{equation}
       \min_{\beta} \| \mathbf{y} - \mathbf{\hat{X}} \boldsymbol{\beta} \|_2^2 + \lambda_2 \| \boldsymbol{\beta} \|_2^2 \quad \text{subject to} \quad \| \boldsymbol{\beta} \|_0 \leq k,
\end{equation}
Here $\beta$ is the vector of weights (scores) assigned to each binary feature, $\lambda_2$ is the $\ell_2$ regularization rate, and $k$ is the model size (bias is omitted for clarity). The sparsity constraint renders the problem NP-hard. We find two methods to train RegScore. First, by relaxing integer constraints on weights in scoring systems, we adapt the BS step from FasterRisk \cite{liu2022fasterrisk} for regression. BS assumes that a model of size $k$ inherently contains one of the best models of size $k-1$. It iteratively expands the solution by optimizing each non-zero weight and selecting the $B$ best candidates. In the second step, BS fine-tunes non-zero weights and retains top-$B$ solutions, ultimately returning the best model. 
The second approach leverages the OKRidge algorithm \cite{NEURIPS2023_80f48ffa}, which is based on the Branch-and-Bound (BnB) algorithm. First, a lower bound is computed for a node in the BnB tree. If this bound is lower than the current solution, a new solution is found and used to generate new nodes in the tree. Both methods yield highly effective RegScore.

\noindent \textbf{Tabular Image Pretraining.} To fully leverage imaging and tabular data, we employ TIP, which integrates a convolutional image encoder $\phi^i$, a transformer-based tabular encoder $\phi^t$, and a multimodal interaction module $\psi$. Given an image representation $I \in \mathbb{R}^{(H'W')\times E}$ from $\phi^i$ and tabular representation $T \in \mathbb{R}^{(N+1)\times E}$ from $\phi^t$ where $E$ denotes the embedding dimension and $N+1$ includes the CLS token, $\psi$ generates a multimodal representation $F \in \mathbb{R}^{(N+1)\times E}$. TIP employs three SSL losses: contrastive learning between $I$ and $T$, image-tabular matching and tabular data reconstruction from $F$. For further implementation details refer to \cite{du2024tip}. Following SSL, we utilize the CLS token during fine-tuning to derive PLR and PRS. 

\noindent \textbf{Personalized Linear Regression (PLR).} In standard transformer architectures, predictions are typically generated by applying a linear layer to the CLS token, mapping it to the number of output classes (1 in the case of regression). Here, we instead transform the CLS token using a linear layer into a vector of size $N+1$, comprising $N$ personalized weights ($\beta_{p,i}$ for each tabular feature $x^t_{p,i}$ per sample $p$) and a bias term ($\beta_{p,0}$). Given these personalized regression weights, the prediction $\hat{y}_p$ is computed using the linear regression equation:
\begin{equation}
       \hat{y_p} = \beta_{p,0} + \sum_{i=1}^{N} x^t_{p,i} \times \beta_{p,i}
\end{equation}

\noindent \textbf{Personalized RegScore (PRS).} PRS follows a similar approach but incorporates binarized tabular features as an additional input. We introduce a gating mechanism that retains only the top $k$ binary features by setting the rest to zero. This is achieved by computing the mean embeddings of each feature and applying a linear transformation ($W_g$) to obtain scores $S$. $k$ features with the highest scores pass through the gating mechanism. During training, we use a soft gating function $K_{s}$ (a sigmoid function with steepness controlled by $\frac{1}{\tau}$), while during inference, we apply a hard gating function $K_{h}$: 
\begin{equation}
    S = W_g\left(\frac{1}{E} \sum_{i=1}^{E} F_{i}\right),\tau_k = \text{topk}(S)_{\text{min}}, K_{\text{h}} = \mathds{1}(S \geq \tau_k), K_{\text{s}}=\sigma\left(\frac{S - \tau_k}{\tau} \right)
\end{equation}
\section{Experiments and results}
\begin{table*}[t!]
    \caption{Results of PH classification with scoring systems of size $k$ and mPAP estimation using RegScore. We \textbf{bold} the best and \underline{underline} second-best results. $\dag \space$ indicates p-value <0.05 for statistically significant difference from RegScore.}
    \label{tab:results_scoring}
    \begin{tabularx}{\textwidth}{lcYYYY}\hline
        \textbf{Method} & $k$ & MAE $\downarrow$ & R $\uparrow$ & Accuracy $\uparrow$ & F1 $\uparrow$  \\ \hline
        Unit \cite{burgess1928factors} & - & - & - & $66.05 \pm 21.5$ & $72.42 \pm 23.0$  \\
        MISS \dag\cite{grzeszczyk2024miss} & 5  & - & - & $84.52 \pm 0.67$ & $90.88 \pm 0.40$ \\
        RiskSLIM \dag\cite{ustun2019learning} & 5 & - & - &  $85.21 \pm 0.62$ & $91.31 \pm 0.47$\\
        FasterRisk \dag\cite{liu2022fasterrisk} & 5  & - & - & $85.36 \pm 1.03$ & $91.40 \pm 0.63$  \\
        FasterRisk \dag \cite{liu2022fasterrisk} & 50  & - & - & $86.51 \pm 1.16$ & $92.04 \pm 0.71$  \\
        \textbf{RegScore$_{BS}$} & 5  & $8.53 \pm 0.15$ & $63.39 \pm 1.44$ & $86.59 \pm 0.27$ & $92.43 \pm 0.15$  \\
        \textbf{RegScore$_{BS}$} & 50  & $\underline{7.75 \pm 0.13}$	& $\underline{69.90 \pm 1.02}$ &	$\underline{88.05 \pm 0.57}$ & $\underline{93.24 \pm 0.31}$  \\
        \textbf{RegScore$_{OKR}$} & 5  &  $8.69 \pm 0.30$ & $61.73 \pm 2.03$ & $86.74 \pm 0.79$ & $92.54 \pm 0.46$ \\
         \textbf{RegScore$_{OKR}$} & 50  &  $\bm{7.73 \pm 0.13}$ & $\bm{70.06 \pm 0.96}$ & $\bm{88.12 \pm 0.72}$ &	$\bm{93.28 \pm 0.38}$ \\
        \hline
    \end{tabularx}
\end{table*}

In what follows, we describe the dataset used for mPAP estimation and PH classification. We compare the performance of RegScore against other methods for constructing classification scoring systems. Additionally, we benchmark PLR and PRS against various tabular and/or image-based approaches.

\noindent \textbf{Dataset.} This study was approved by the Ethics Committee. The dataset originates from the ASPIRE Registry (Assessing the Severity of Pulmonary Hypertension In a Pulmonary Hypertension REferral Centre) \cite{Hurdman2012} and comprises 2051 invasively measured mPAP values matched with Cardiac MRI (CMR) videos of one cardiac cycle (short-axis plane). It includes data from 1918 patients (1171 females, 747 males, aged $64 \pm 14$ years) with some undergoing repeated procedures over time. We select demographic features and MRI-derived measurements with fewer than 500 missing values. The CMRs were acquired using devices from multiple vendors including Siemens, Philips and GE. Instead of using full videos, we extract systolic, diastolic and in-between frames as 3-channel images \cite{hager2023best,du2024tip}.

\begin{figure}[t!]
    \centering
    \includegraphics[width=\textwidth]{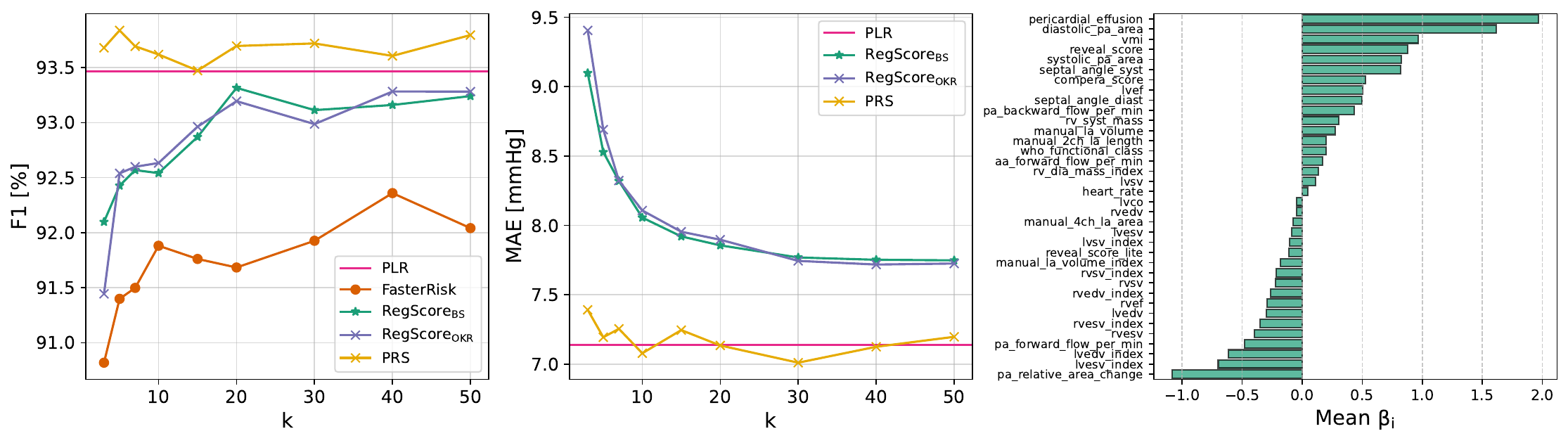}
    \caption{Performance comparison of classification (left) and regression (center) across model sizes ($k$). The right panel shows the mean feature weights in PLR.}
    \label{fig:model_size}
\end{figure}

\noindent \textbf{Implementation details.} We split the dataset into the training set (1790 samples)  used for a 5-fold cross-validation and test set (261 samples), ensuring that each patient's data appears in only one split. The splits are stratified based on mPAP (divided into four bins) to maintain similar distributions. For the scoring systems' classification task, cases with mPAP exceeding 25 mmHg are considered positive (1678 positive vs. 373 negative cases). For other methods trained on regression task, mPAP serves as ground truth and classification is achieved by thresholding the predicted value. We use Mean Absolute Error (MAE) and Pearson's correlation coefficient (R) as regression metrics while accuracy and F1 as classification metrics. Mean and standard deviation are reported across the test set over five folds. We standardize numerical features, retaining only those with statistical significance based on f-regression \cite{scikit-learn}. We present all features as part of Fig. \ref{fig:model_size}. CMRs are resampled to a pixel spacing of 0.9375mm$\times$0.9735mm and resized to 128$\times$128 pixels. Deep learning models are implemented in PyTorch and trained on an NVIDIA A100 80GB GPU for up to 500 SSL and fine-tuning epochs with the Adam optimizer. The best model is selected by validation performance, with SSL and fine-tuning learning rates chosen from \{$3\times10^{-3}$, $3\times10^{-4}$, $3\times10^{-5}$\} and \{$1\times10^{-3}$, $5\times10^{-4}$, $1\times10^{-4}$\} respectively. We set $B=10$, $\tau=0.1$, $k=5$, $\lambda_2=10^{-8}$, and use tertiles for $\xi$ in PRS and MDLP in RegScore.

\begin{table*}[t!]
    \caption{Results of mPAP regression and PH classification using imaging (I) and/or tabular (T) methods with SSL and supervised learning (FT). $\dag \space$ indicates p-value <0.05 between the performance of PLR, PRS and other methods.}
    \label{tab:results}
    \begin{tabularx}{\textwidth}{lccYYYY}\hline
        \textbf{Method} & SSL & FT & MAE $\downarrow$ & R $\uparrow$ & Accuracy $\uparrow$ & F1 $\uparrow$  \\ \hline

        \multicolumn{7}{c}{\textbf{Machine Learning methods}} \\ \hline
        DT \dag \cite{scikit-learn} & - &  T& $10.97 \pm 0.29$  & $51.43 \pm 2.80$  & $81.23 \pm 1.46$ &	$88.74 \pm 0.96$ \\
        XGB \dag \cite{xgboost} &  - & T & $8.05 \pm 0.07$ & $69.27 \pm 1.23$ & $87.36 \pm 0.61$ &	$92.77 \pm 0.35$ \\
        LR \dag \cite{scikit-learn} & - & T & $7.76 \pm 0.06$	& $69.66 \pm 0.79$ & $87.89 \pm 0.34$ & $93.19 \pm 0.18$\\
        GBR \dag \cite{scikit-learn} &  - & T & $7.67 \pm 0.06$ & $70.80 \pm 0.63$ & $87.89 \pm 0.21$ &	$93.13 \pm 0.11$\\
        RF \dag \cite{scikit-learn} & - & T & $7.61 \pm 0.04$  & $70.42 \pm 0.68$ & $87.13 \pm 0.88$ &	$92.69 \pm 0.53$ \\\hline
        
        \multicolumn{7}{c}{\textbf{Deep Learning methods}} \\ 
        \hline
        ResNet-50 \dag \cite{he2016deep} & - & I & $8.24 \pm 0.28$ & $68.25 \pm 2.74$ & $84.83 \pm 1.62$ & $91.55 \pm 0.67$ \\  
        SimCLR \dag \cite{chen2020simple} & I & I & $7.91 \pm 0.30$ & $72.25 \pm 2.12$ & $85.59 \pm 0.75$ & $91.76 \pm 0.33$ \\
        DAFT \dag \cite{polsterl2021combining} & - & IT  & $7.74 \pm 0.30$ & $71.24 \pm 1.02$ & $88.51 \pm 1.27$ & $93.29 \pm 0.80$ \\  
        TabMixer$_{2D}$ \dag \cite{grzeszczyk2024tabmixer} & - & IT  & $7.69 \pm 0.14$ & $72.08 \pm 0.44$ & $88.35 \pm 0.79$ & $93.23 \pm 0.41$ \\ 
        VIME \dag \cite{vime} & T & T & $7.53 \pm 0.09$ & $73.22 \pm 0.88$ & $\underline{88.97 \pm 1.25}$ & $93.56 \pm 0.73$ \\
        TabAttention$_{2D}$    \dag   \cite{grzeszczyk2023tabattention} & - & IT & $7.47 \pm 0.11$ & $72.83 \pm 1.14$ & $88.89 \pm 0.72$ & $\underline{93.59 \pm 0.31}$ \\    
        MMCL \cite{hager2023best} & IT & I & $7.45 \pm 0.37$ & $75.04 \pm 1.60$ & $87.89 \pm 0.44$ & $92.93 \pm 0.25$ \\ 
        TIP \cite{du2024tip} & IT & IT & $\bm{6.88 \pm 0.25}$ & $\bm{77.30 \pm 1.46}$ & $88.58 \pm 0.63$ & $93.39 \pm 0.35$\\
         \textbf{PLR} & IT & IT & $\underline{7.14 \pm 0.14}$ & $\underline{75.07 \pm 0.87}$ & $88.66 \pm 1.29$ & $93.46 \pm 0.73$\\
        \textbf{PRS$_{5}$} & IT & IT &  $7.19 \pm 0.16$ &$ 74.85 \pm 1.26$ & $\bm{89.43 \pm 1.00}$ & $\bm{93.84 \pm 0.60}$   \\
        \hline
    \end{tabularx}
\end{table*}

\noindent \textbf{Comparison with state-of-the-art methods.} We compare the performance of RegScore against other scoring system methods, including Unit \cite{burgess1928factors}, MISS \cite{grzeszczyk2024miss}, RiskSLIM \cite{ustun2019learning} and FasterRisk \cite{liu2022fasterrisk}. The results of these experiments are presented in Table \ref{tab:results_scoring}. Both RegScore training approaches outperform competing methods on classification metrics (with statistically significant differences, paired t--test p--value <0.05) for $k=5$ and $k=50$, while also providing interpretable mPAP estimation. RegScore trained with OKRidge achieves slightly better performance than the BS version, however, the difference is not statistically significant. We present examples of scoring systems in Table \ref{tab:example_scoring_systems}. We also compare PLR and PRS against machine learning models trained on tabular data, including LR, XGBoost \cite{xgboost}, Gradient Boosting Decision Trees (GBDT), and Random Forest (RF). Additionally, we benchmark them against deep learning methods for imaging and/or tabular data, including ResNet-50 \cite{he2016deep}, SimCLR \cite{chen2020simple}, DAFT \cite{polsterl2021combining}, TabMixer \cite{grzeszczyk2024tabmixer}, VIME \cite{vime}, TabAttention \cite{grzeszczyk2023tabattention}, MMCL \cite{hager2023best}, and TIP \cite{du2024tip} (Table \ref{tab:results}). Although PLR and PRS yield slightly higher MAE values, 7.14 and 7.19, respectively, compared to TIP (6.88), they outperform all other methods in regression performance, with all but one difference being statistically significant. Notably, PRS achieves the highest classification metrics (F1 = 93.84) among the evaluated approaches, while also offering interpretability.

\noindent \textbf{Ablation study.} We conduct an ablation study (Table \ref{tab:ablation}) to assess key aspects of our methods. The performance of both PLR and PRS worsens when trained without image data or SSL, highlighting the importance of the training procedure and bimodality. In all methods, modifying the discretization function leads to a decline in performance, underscoring the need to carefully select an appropriate binning strategy for the algorithm.

\begin{table*}[t!]
    \caption{Ablation study of the key components in proposed methods.}
    \label{tab:ablation}
    \begin{tabularx}{\textwidth}{lYYYY}\hline
        \textbf{Method} & MAE $\downarrow$ & R $\uparrow$ & Acc. $\uparrow$ & F1 $\uparrow$  \\ \hline
        \textbf{RegScore}$_{BS5}$ & $\bm{8.53 \pm 0.15}$ & $\bm{63.39 \pm 1.44}$ & $\bm{86.59 \pm 0.27}$ & $\bm{92.43 \pm 0.15}$  \\
        $\>$ w/ tertile bins & $9.38 \pm 0.36$ &	$53.47 \pm 3.54$ &	$86.28 \pm 1.81$ &	$92.28 \pm 0.89$  \\ \hline
        
        \textbf{RegScore}$_{OKR5}$ & $\bm{8.69 \pm 0.30}$ & $\bm{61.73 \pm 2.03}$ & $\bm{86.74 \pm 0.79}$ & $\bm{92.54 \pm 0.46}$ \\
        $\>$ w/ tertile bins & $9.41 \pm 0.27$ &	$52.50 \pm 3.36$ &	$86.21 \pm 1.51$ &	$92.24 \pm 0.81$  \\
        \hline

        \textbf{PLR}  &  $\bm{7.14 \pm 0.14}$ & $\bm{75.07 \pm 0.87}$ & $\bm{88.66 \pm 1.29}$ & $\bm{93.46 \pm 0.73}$ \\
        $\>$ w/o Image & $8.78 \pm 0.91$ & $62.28 \pm 9.13$ & $86.59 \pm 2.12$ & $92.41 \pm 1.13$   \\
        $\>$ w/o SSL & $7.79 \pm 0.29$ & $70.00 \pm 1.56$ & $88.28 \pm 0.58$ & $93.32 \pm 0.30$ \\ \hline
        
        \textbf{PRS$_{5}$} & $\bm{7.19 \pm 0.16}$ & $\bm{74.85 \pm 1.26}$ & $\bm{89.43 \pm 1.0}$ & $\bm{93.84 \pm 0.6}$   \\
        $\>$ w/ MDLP bins & $7.19 \pm 0.33$ & $74.35 \pm 2.29$ & $88.74 \pm 0.34$ & $93.50 \pm 0.21$\\
        $\>$ w/o SSL & $7.93 \pm 0.22$ & $70.10 \pm 3.00$ & $88.97 \pm 0.32$ & $93.58 \pm 0.17$ \\ 
        $\>$ w/o Image & $7.60 \pm 0.13$ & $71.53 \pm 1.00$ & $88.97 \pm 0.50$ & $93.60 \pm 0.24$  \\ \hline
    \end{tabularx}
\end{table*}

\begin{table}[t!]
\caption{Examples of RegScore for mPAP estimation and other scoring systems for PH classification.}
\label{tab:example_scoring_systems}

\centering
    \begin{subtable}[t]{.4\textwidth}
    \caption{\textbf{RegScore}}
    \adjustbox{width =\textwidth}{\begin{tabular}{lr}\hline
        \textbf{feature} & \textbf{mPAP}\\\hline
        $\bm{bias}$ & 40.46\\
        $\bm{1.0 \leq reveal\_score < 6.0}$ & -11.23\\
        $\bm{0.73 \leq vmi \leq 2.87}$ & 8.14\\
        $\bm{49.17 \leq rvef \leq 82.0 }$ & -5.90\\
        $\bm{311.46 \leq systolic\_pa\_area < 672.0}$ & -8.52\\
        $\bm{166.0 \leq septal\_angle\_syst}$ & 10.71\\
        \hline
        \textbf{mPAP [mmHg]:} & = $\sum$ \\\hline
        \end{tabular}
    }
    \end{subtable}
    \begin{subtable}[t]{.365\textwidth}
        \caption{FasterRisk}
        \adjustbox{width =\textwidth}{\begin{tabular}{lr}\hline
            \textbf{feature} & \textbf{points}\\\hline
            $\bm{bias}$ & 4\\
            $\bm{1.0 \leq reveal\_score < 6.0}$ & -3\\
            $\bm{72.9 \leq rv\_syst\_mass \leq 259.04}$ & 3\\
            $\bm{270 \leq diastolic\_pa\_area < 545}$ & -3\\
            $\bm{166 \leq septal\_angle\_syst}$ & 4\\
            $\bm{pericardial\_effusion = No}$ & -2\\
            \hline
            \multicolumn{2}{c}{\textbf{risk PH:} $1/(1 +  exp(-score))$}\\\hline
            \end{tabular}
            }
    \end{subtable}
    \begin{subtable}[t]{.365\textwidth}
    \caption{RiskSLIM}
    \adjustbox{width =\textwidth}{\begin{tabular}{lr}\hline
        \textbf{feature} & \textbf{points}\\\hline
        $\bm{bias}$ & 3\\
        $\bm{1.0 \leq reveal\_score < 6.0}$ & -2\\
        $\bm{rv\_dia\_mass\_index < 12.56}$ & -1\\
        $\bm{270 \leq diastolic\_pa\_area < 545}$ & -2\\
        $\bm{166 \leq septal\_angle\_syst}$ & 2\\
        $\bm{pericardial\_effusion=No}$ & -1\\
        \hline
        \multicolumn{2}{c}{\textbf{risk PH:} $1/(1 +  exp(-score))$}\\\hline
        \end{tabular}
    }
    \end{subtable}
    \begin{subtable}[t]{0.40\textwidth}
        \caption{MISS}
        \centering
        \adjustbox{width =\textwidth}{
        \begin{tabular}{l*{2}{r}}\hline
                \textbf{feature} &  \textbf{No PH}& \textbf{PH}\\\hline
            $\bm{bias}$& -5& -4\\
            $\bm{1.0 \leq reveal\_score < 6.0}$& 1& -1\\
            $\bm{12.56 \leq rv\_dia\_mass\_index}$& -4& -3\\
            $\bm{270 \leq diastolic\_pa\_area < 545}$& 0& -2\\
            $\bm{166 \leq septal\_angle\_syst}$& -3& -1\\
            $\bm{pericardial\_effusion=UNK}$& -1& 0\\
            \hline
            \textbf{score:}& = $\sum$& = $\sum$\\\hline
        \end{tabular}
        }
    \end{subtable}

\end{table}

\section{Discussion and Conclusions}
In this paper, we introduced scoring systems for regression tasks. For diseases like PH which are diagnosed by thresholding specific measurements, RegScore offers higher clinical interpretability. Unlike traditional scoring systems, RegScore not only provides an interpretable prediction but also directly relates it to the measure of interest. RegScore outperformed other scoring systems on the classification task (p--value <0.05). 
RegScore is efficient to calculate - it can be generated in minutes ($\approx$55 seconds for BS variant), whereas deep learning models require hours for training. This speed enables the exploration of multiple near-optimal models to select the best one by domain experts. Furthermore, PLR and PRS introduce almost no computational overhead compared to TIP, as they only require a linear layer with additional parameters ($N$ or $D$ outputs instead of 1). 

We examined the impact of model size $k$ on the performance of RegScore and PRS, with results presented in Fig. \ref{fig:model_size}. Across all model sizes, RegScore achieves better classification results than FasterRisk, with larger models yielding improved classification and regression performance. PRS results remain stable due to the model’s personalized nature, which adapts to the number of selected features. Because PLR directly couples tabular features to predictions, we can analyze feature importance. In Fig. \ref{fig:model_size}, we present mean weight values for each feature in PLR. High coefficients are assigned to features also selected by RegScore (e.g. systolic septal angle, reveal score), aligning with findings from other studies \cite{lungu2016diagnosis}. 

Our methods have limitations. There is a trade-off between interpretability and performance. Shifting from the most effective black-box TIP toward RegScore increases interpretability but reduces regression performance. This trade-off arises because PLR, PRS, and RegScore constrain their predictions by coupling them with tabular data. However, this flexibility allows clinicians to choose between more interpretable or higher-performing models based on their needs. Future work could address this trade-off by incorporating binning into the optimization process to enhance performance while maintaining interpretability.

In summary, we introduced RegScore, a novel approach for interpretable scoring in regression tasks, along with PLR and PRS, which enhance interpretability in bimodal models. Our results show that RegScore outperforms existing scoring systems in PH classification and holds promise for broader clinical applications.

\begin{credits}
\subsubsection{\ackname} This research has been made possible by the Kosciuszko Foundation. The American Centre of Polish Culture. This research was funded in whole or in part by National Science Centre, Poland 2023/49/N/ST6/01841. For the purpose of Open Access, the author has applied a CC-BY public copyright licence to any Author Accepted Manuscript (AAM) version arising from this submission. This work is supported by the EU’s Horizon 2020 programme (grant no. 857533, Sano) and the Foundation for Polish Science’s International Research Agendas programme (MAB PLUS/2019/13), co-financed by the EU under the European Regional Development Fund and the Polish Ministry of Science and Higher Education (contract no. MEiN/2023/DIR/3796).

\subsubsection{\discintname}
The authors have no competing interests to declare.
\end{credits}
%
%
%
%

\bibliographystyle{splncs04}
\bibliography{bibliography}

\end{document}